\documentclass[12pt]{article}
\usepackage{aaspp4,flushrt}
\lefthead{Lehnert et al.}
\righthead{HST Snapshot Survey of 3CR Quasars}

\begin{document}

\title{\bf HST Snapshot Survey of 3CR Quasars: The Data
\altaffilmark{1}}

\bigskip
\bigskip
\author{Matthew D. Lehnert\altaffilmark{2,3},
George K. Miley\altaffilmark{2},
William B.  Sparks\altaffilmark{4},
Stefi A. Baum\altaffilmark{4},
John Biretta\altaffilmark{4},
Daniel Golombek\altaffilmark{4},
Sigrid de Koff\altaffilmark{2,4},
Ferdinando D. Macchetto\altaffilmark{4,5}, and
Patrick J. McCarthy\altaffilmark{6}}

\medskip

\altaffiltext{1}{Based on observations with
the NASA/ESA {\it Hubble Space Telescope}, obtained at the Space
Telescope Science Institute, which is operated by the Association
of Universities for Research in Astronomy, Inc., under NASA contract
NAS5-26555.}
\altaffiltext{2}{Sterrewacht Leiden, Postbus
9513, 2300 RA Leiden, The Netherlands}
\altaffiltext{3}{Current address: Max-Planck-Institut f\"ur extraterrestrische
Physik (MPE), Postfach 1603, D-85740 Garching, Germany, mlehnert@mpe.mpg.de}
\altaffiltext{4}{STScI, 3700 San Martin Dr., Baltimore, MD 21218}
\altaffiltext{5}{Affiliated with the Space Science Department, ESA}
\altaffiltext{6}{Observatories of the Carnegie Institution,
Pasadena, CA}

\medskip

\vskip 0.1in
\centerline{Received ........................; accepted ........................
}

\newpage

\begin{abstract}

We present images taken with the Wide Field Planetary Camera (WFPC-2)
on the Hubble Space Telescope of 43 quasars selected from the 3CR radio
catalog.  The redshift range of the targets is large --- 0.3 $\lesssim$ z
$\lesssim$ 2 and allows us to probe the nature of quasar hosts from about
20\% to 80\% of the age of the universe.  These data were taken in the
course of a large program that imaged 267 3CR radio galaxies and
quasars using the HST in ``snapshot'' mode. Each quasar was centered on
the Planetary Camera (PC1) and was imaged through the F702W filter
(bandpass similar to Cousins R).  Typical integration times were 5 and 10
minutes.  For each quasar, we attempted to judge the contribution of
the host galaxy to the total light from the quasar in two ways.  The
first method was to compare the radial light distributions of the
quasars with that of both model point spread function and an empirical
PSF constructed by summing individual observations of standard stars.
Second, to provide morphological information we attempted to remove the
contribution of the quasar nucleus from the extended emission by
subtracting a point-spread-function constructed from observations of
standard stars.  This second method proved to be more sensitive in detecting
marginally extended emission.

Our analysis suggests that the quasar fuzz contributes from $<$5\% to
nearly 100\% in the most extreme case (about 20\% being typical) of the
total light from the quasar, with 16 of the quasars ($\sim$40\%) being
unresolved according to the analysis of their light profiles (with only
7 being considered unresolved determined by PSF subtraction of the
quasar images).  The magnitudes of the hosts range from about 18 to
$>$21 in the F702W filter and the sizes are typically 1 to 2 arc
seconds at a limiting surface brightness of $\sim$21--22 m$_{F702W}$
arcsec$^{-2}$.  Comparisons with the few ground-based images that are
available of these sources suggest good overall morphological agreement
with the HST images.  The 0.1'' resolution of the HST PC combination
reveals a wide variety of structures in the host galaxies of these
quasars.  Most of the host galaxies show twisted, asymmetric, or
distorted isophotes.  About 1/4 of the quasar hosts have close (within
a few arc seconds) companions seen in projection and about 1/10 show
obvious signs of tidal interactions with a close companion.  Finally,
using radio images available from the literature, we find that in many
of the resolved sources there is a correspondence between the radio and
optical morphologies.  We find that these sources exhibit a tendency for
the principal axes of the radio and optical emission to align similar
but perhaps weaker than that observed for radio galaxies.  This
correspondence also suggests that our methodology for removing the
point source contribution from the resolved emission is sound.  A
more complete analysis of these data and new HST snapshot data will be
presented in subsequent papers.

\end{abstract}


\newpage

\section{Introduction}

Observations of radio-loud quasars are important for investigating some
of the most interesting and fundamental problems of contemporary
astrophysics. The foremost of these is the investigation of causes of
the ``extinction'' of luminous quasars.  The space density of luminous
quasars has declined by about 3 orders-of magnitude from the epoch
z=2--3 to the present (e.g., Hartwick \& Schade 1989).  What processes
led to the such a dramatic decrease in co-moving space density?  Are
these processes related to their galactic and/or cluster scale
environments?  The answers to these questions will be important in
furthering our understanding galaxy evolution.

Studies of the environments of quasars can also provide insight into
the AGN phenomenon in general.  Of contemporary interest is the
relationship between quasars and radio galaxies.  A variety of schemes
have been proposed to link quasars and radio galaxies through
differences in their environments, viewing angle, or evolutionary state
(e.g., Norman \& Miley 1984; Barthel 1989; Neff \& Hutchings 1990).  In
particular, the \lq\lq viewing angle\rq\rq \ scheme of Barthel (in
which radio-loud quasars and radio galaxies are drawn from the same
parent population but viewed preferentially at small or large angles,
respectively, to the radio axis) predicts that the luminosity and color
of the quasar fuzz should be identical to those of the radio galaxies
at similar radio powers and redshifts (as do some of the evolutionary
unification schemes).  Also, radio galaxies at high redshifts (z
$\gtrsim$ 0.6) exhibit the so-called \lq\lq alignment effect\rq\rq
\ (McCarthy et al. 1987; Chambers et al. 1987) namely that the radio, and
the rest-frame UV and optical axes are all roughly co-linear.  If
indeed quasars and radio galaxies are objects differentiated only by
viewing angle, then quasars might also be expected to exhibit such an
alignment effect over the same redshift range as the radio galaxies.

Investigating the above problems with ground-based data has been
hampered by the inability of separating marginally extended ``fuzz''
from the ``blinding'' light of the quasar nucleus.  But despite these
difficulties, limited information about the properties of the host
galaxies of quasars spanning a range of redshifts and radio properties
has been obtained (see e.g., Hutchings, Crampton, \& Campbell 1984; Smith
et al. 1986; Stockton \& MacKenty 1987; Veron-Cetty \& Woltjer 1990;
Heckman et al. 1991).

The Hubble Space Telescope is well-suited for investigating quasar host
galaxies.  Because of its high spatial resolution, imaging with the HST
allows us to remove the contribution of the quasar nucleus and
investigate the properties of the host galaxies on scales of 0.1 to a
few kpc, depending on the redshift of the source.  This was one of the
reasons that motivated us to undertake a ``snapshot survey'' of sources
in the 3CR catalogue.  The ``snapshot'' mode of observing, in which
gaps in the primary HST schedule are filled in with short integrations
of selected targets, greatly enhances the overall efficiency of the HST
and is well-suited to observing large samples of objects.  The results
presented here should be compared and contrasted with recent HST
results by other groups.  Images of low/modest redshift quasars
obtained with the WFPC-2 on HST show a wide variety of host galaxy
morphologies including elliptical and spiral systems as well as
highly disturbed, galaxies which may be interacting with and/or
accreting close companions (Bahcall et al. 1995a,b; Disney et al. 1995;
Hutchings \& Morris 1995; Hooper et al. 1997).

The 3CR data set allows the properties of matched samples of radio-loud
quasars and radio galaxies to be compared and should ultimately allow
us to investigate the properties of ``fuzz'' over a wide range of
redshifts (0.3 $\lesssim$ z $\lesssim$ 2), radio luminosities (log P$_{178
MHz}$ $\sim$ 10$^{27}$ to 10$^{29.3}$ W Hz$^{-1}$) and radio types (e.g.,
lobe-dominated to core-dominated).  Since low-frequency radio emission
from radio-loud AGN is thought to be emitted isotropically, the
low-frequency selection of the 3CR (178 MHz) implies that it is
relatively unbiased by anisotropic emission (whether it is due to
relativistic beaming, emission from an optically thick accretion disk,
or due to an obscuring torus), thus the 3CR is particularly well suited
for investigating the relationship between radio galaxies and quasars.
Here we present and describe the image analysis and data reduction for
43 quasars from this ``3CR snapshot survey''.  Other papers in this
series have described the properties of the radio galaxies (de Koff
et al. 1996; Baum et al. 1999; McCarthy et al. 1997) and future work will
compare matched samples of radio galaxies and quasars over a wide range
of redshifts.

\section{Sample Selection}

To select objects for the snapshot survey, we used the revised 3CR
sample as defined by Bennett (1962a,b), having selection constraints of
(i) flux density at 178 MHz, S(178) $>$ 9 Jy, (ii) declination $>$
5$^\circ$, and (iii) galactic latitude, $|b| >$ 10$^\circ$.  All of the
sources have been optically identified and have measured redshifts
(Spinrad et al. 1985; Djorgovski et al. 1988).  In this paper, we report
on the properties of the quasars observed during the snapshot survey of
a total of 267 3CR sources.  The classification of these sources as
quasars are based on the compilations of Spinrad et al. (1985) and
Djorgovski et al. (1988).  Of the total of 53 quasars listed in Spinrad
et al. (1985), we have observed 41 (3C 179 and 3C 279, which we
have observed, were not listed in the Spinrad et al. quasar
identifications).  Inevitable scheduling constraints and conflicts with
other observing programs accounts for the 12 3CR quasars that were not
imaged as a part of this program.  Table 1 gives a summary of the
observations.

\section{Characteristics of the Observed Sample}

The characteristics of the observed sample of quasars are provided in
Table 2 and summarized graphically in Figures 1 through 3.  The
redshift distribution of the observed sample is fairly flat and ranges
from 0.3 to about 2.1 (Table 2 and Fig. 1) and is broadly similar to
that of the entire sample of 3CR quasars, the major difference is that
a few of the lowest redshift quasars are excluded.

Adopting a cosmology with a Hubble constant of H$_0$ = 75 km s$^{-1}$
Mpc$^{-1}$ and a deceleration parameter of q$_0$=0.5 and using the 178
MHz radio fluxes of the sample as listed in Spinrad et al. (1985), the
radio (178 MHz) power distribution of the observed sample lies in the
range log P$_{178 MHz}$ = 27 --- 29.3 W Hz$^{-1}$.  The
distribution has a strong peak at log P$_{178 MHz}$ $\approx$ 28.5
W Hz$^{-1}$ (Figure 2 and Table 2).  Hence the observed quasars are
amongst the most powerful known radio sources.

Within the sample, the radio emission from these sources also exhibit a
wide variety of physical sizes.  Again, adopting the cosmology of H$_0$
= 75 km s$^{-1}$ Mpc$^{-1}$ and q$_0$=0.5 and measuring the largest
angular size of the radio sources using radio maps available in the
literature (see Table 2 and \S 7), shows that these sources range from
compact, galaxy-sized radio sources (largest projected linear sizes
$\lesssim$10 kpc) to cluster-scale sources with sizes of a few hundred
kpc (Figure 2).

\section{Observations and Pipeline Reduction}

The observations were taken throughout HST Cycle 4 (early 1994 to mid
1995) and the integrations times ranged from 2 to 10 minutes, with 5 or
10 minutes with being typical (see Table 1).  All the quasars were
observed close to the center of the Planetary Camera (PC) and imaged
through the F702W filter, whose bandpass corresponds approximately to
that of the Cousins R filter.  The camera/filter combination was chosen
to give the maximum spatial and photometric sensitivity.  The large
width of the F702W filter bandpass means that every quasar image has
some contribution to its brightness and morphology due to one or two
prominent emission lines.  From the shape of the system bandpass for
the F702W filter available in the HST WFPC2 Instrument Handbook
(1995), the prominent lines that contribute to the
emission within the images and the redshift range over which they
contribute are:  [OIII]$\lambda$5007 (0.19 $<$ z $<$ 0.64), H$\beta$
(0.22 $<$ z $<$ 0.67), [OII]$\lambda$3727 (0.60 $<$ z $<$ 1.20),
MgII$\lambda$2798 (1.13 $<$ z $<$ 1.93).  Only 3C 9 has a redshift
(z=2.012) that avoids having prominent emission lines within the
bandpass of the F702W filter.

The images were reduced using the standard pipeline (see the HST Data
Handbook, Version 2, 1995).  The standard pipeline
includes bias subtraction, dark count correction, flat-field
correction, and a determination of the absolute sensitivity.  For those
objects that had two or more individual exposures (see Table 1), the
separate images were combined using the STSDAS task CRREJ.  This task
constructs an average of the input frames and iteratively removes
highly deviant pixels from the average.  For those quasars with only
one exposure, we used the IRAF task ``cosmicrays'' to remove the
effects of cosmic rays.  This task detects pixels that have
significantly different value than the surrounding pixels and replaces
the deviant value with the average of the surrounding pixels.  Weak
cosmic rays that were missed using this technique were subsequently
removed by fitting the background to the area immediately surrounding
the suspected cosmic ray hit and using this new value as a substitute
for the old value of the counts in the affected pixel.

The data were flux-calibrated using the inverse sensitivity for the
F702W filter of 1.834 $\times$ 10$^{-18}$ ergs s$^{-1}$ cm$^{-2}$
\AA$^{-1}$ dn$^{-1}$ and a zero point of 22.469 (Whitmore 1995).  This
puts the magnitudes on the ``Vega system''.  To convert to the STMAG
system, which assumes a flat spectral energy distribution and has a
zero point of 23.231, 0.762 magnitudes should be subtracted from the
magnitudes given here.

\section{Image Reduction}

\subsection{PSFs and EEDs}

One of the most important steps of the analysis is to quantify the
amount of extended emission from these quasar images.  We have
attempted this using two methods.  To do this we collected images of
the standard stars used to calibrate the F702W filter that were near
the dates of the observations.  Unfortunately, there were only a few
exposures (5) that were useful.  These were then used in two ways.  We
constructed an empirical point-spread function (PSF) using these
observations by adding up the individual exposures after they had been
aligned to a common center.  This empirical PSF was then compared with
the model PSF constructed using the PSF modeling program, Tiny Tim.
However, as we discuss below, the close agreement is limited to
azimuthal averages -- Tiny Tim does not reproduce the detailed
two-dimensional structure of the PSF.  Next, we measured the encircled
energy diagrams (EEDs), defined as the fraction of the flux from a
point source interior to a radius r, as a function of r. We then
intercompared all the EEDs (including the PSF generated by Tiny Tim)
taken through a given filter to determine the reproducibility of the
EED.  There was very good agreement between the shape of the EED from
the sum of the observations of standard stars and that of the Tiny Tim
PSF but insufficient data were available to perform this comparison for
individual stars/PSFs taken with the F702W filter.  Fortunately, we
also carried out this analysis for another of the WFPC2 filters, F555W,
in the course of another HST program (Lehnert et al. 1999).  For this
intercomparison, we used observations of approximately 20 stars.  This
intercomparison implies that we can robustly detect fuzz that
contributes more than about 5\% as much light as the quasar itself
(within a radius of about 1.4 arcsec).  This limit is consistent with
the known temporal variations in the HST PSF due to effects like the
gentle change in focus over times scales of months and shorter
(orbital) time scale variations due to the so-called ``breathing'' of
the telescope (see WFPC-2 Instrument Handbook).  We have restricted ourselves
to radii less than about 1.5 arc seconds due to the poorly understood
large angle scattering which becomes important beyond a radius of about
2''.

To quantify how much of the quasar light is extended, we compared the
EEDs of each quasar to that of the PSF (both empirical and model,
although it makes little difference which one is used).  We
accomplished this by scaling the PSF EED so that the total flux
difference between the PSF EED and that of each quasar is zero inside a
radius of 2 PC pixels ($\approx$0.09'').  The EED analysis is
conservative in that the actual underlying fuzz will certainly have a
finite non-zero central surface brightness.  Therefore, this analysis
provides a lower limit to that amount of extended flux coming from each
quasar.  Moreover, the choice of using a radius of 2 pixels was not
arbitrary.  Experimenting with the effect of scaling by the central
pixel led to a much higher variation in the structure of each stellar
EED compared to the EED of the model PSF.  This variation is due to our
limited ability in determining the exact location of the peak of
emission and also due to the fact that the central point source never
falls exactly in the center of one pixel.  Enlarging the area overwhich
the quasar and PSF are scaled greatly reduced the variation in the
structure seen in the PSF by the EED analysis.  However, selecting a
large area leads to very limited sensitivity to extended emission and
thus a radius of 2 pixels was chosen as the most suitable choice to
maintain the sensitivity to mildly extended fuzz but also minimizing
systematic problems.  In addition, we limited the EED analysis to
a radius $\leq$30 PC pixels ($\approx$1.4'').  This mimizes the
problem of the large angle scattering which may not be well represented
in the empirical or model PSF but may contribute to the extended
emission in the quasar images.  Therefore, in Table 3, we quote the
amount of extended emission as a fraction of the total quasar emission
within a radius of 30 PC pixels (1.4'').  In Figure 4, we show all of
the profiles from the EED analysis.

The EED procedure assumes that the peak flux of the underlying galaxy
is zero.  Of course we know that this is a lower limit.  To investigate
how much flux we may miss in the course of such a procedure, we
conducted the EED analysis for a small sample of 10 radio galaxies also
observed as part of the snapshot survey.  These galaxies were chosen to
represent the full range of morphologies and (most importantly) central
surface brightnesses seen in the data on radio galaxies (see de Koff
et al. 1996; McCarthy et al. 1997).  Although the images reveal
non-thermal nuclei in a few objects, none are as ``PSF dominated'' as
any of the quasars.  Under the assumption that radio galaxies are
similar to quasar hosts in morphology and central surface brightness
and that the central surface brightness observed are due to the
underlying galaxies and not the AGNs, we determined the EED fluxes for
the radio galaxies. These fluxes were smaller than the measured total
fluxes by an amounts ranging from about 8\% to 50\% (with about 30\%
being the average) of the total extended light.  We regard this as a
reasonable underestimate of the quasar fluxes by the EED analysis.

\subsection{PSF Subtraction}

There are obvious limitations to using the EED analysis described
above.  The foremost of these is that it does not provide information
on the morphology of the host galaxy.  Also, it in fact only provides a
lower limit on the amount of extended emission since we have assumed
that the contribution to the fuzz from the central 0.03 $\sq\arcsec$ is
zero.  To study the host galaxy morphology, it is necessary to first
remove the contribution of the quasar nuclei from the images.
Therefore to provide morphological information about the host galaxies,
we next attempted subtraction of a scaled PSF from each of the quasar
images.  We used two different PSFs, a model of the HST PSF constructed
using Tiny Tim and an empirical PSF constructed by averaging several
images of standard stars.  We would have preferred to construct a PSF
using exposures of open clusters or of outer regions of globular
clusters but no such images were available from Cycle 4 observations
taken through the F702W filter. Also only a limited number
of exposures of standard stars were available and some of the stellar images
were far (up to about 300 pixels) from the center of the PC.  If there
were two or more images of the same standard taken at the same position
close in time, the individual exposures were averaged. Images of
different standards were then summed after being aligned to the nearest
pixel. The model and empirical PSFs were scaled such that their peaks
were about 5 - 15\% of the highest valued pixel in the quasar image and
iteratively subtracted until emission due to the diffraction of the
secondary support become negligible or the flux in the central pixel of
the quasar image was too small to be measured.  During this procedure
we noted the subtraction level where residual diffraction spikes were
just above the background noise level.  Also, we continued the
subtraction until a negative image of the diffraction spikes appeared
above the level of the noise.  This procedure allowed us to estimate
the uncertainty in the fraction of extended emission by observing the
values where the diffraction spikes in the residual image became
negative due to over-subtraction of the PSF or were still present due
to under-subtraction.

The relatively subjective method described above allows us to
parameterize the uncertainty in the PSF subtraction procedure as a
function of the total brightness of the quasar.  We found that for most
of the quasars, the uncertainty in the flux of the host relative to
that of the total (quasar plus host) was about $\pm$7\%.  For about
20\% of the sample (8 quasars), the uncertainty was larger, about
$\pm$15\% of the total quasar flux.  We used these estimates to
categorize the quasars into three groups according to the uncertainties
in the fluxes of the remnant hosts.  The three categories correspond to
$\sim$$\pm$0.2, $\sim$$\pm$0.4, and $\sim$$\pm$0.7 magnitudes.

After conducting the subtraction process with both the model PSF and
the empirical PSF for about 10 of the quasar images, we concluded that
the model PSF was inadequate for PSF subtraction.  There is an
asymmetry in the intensity of the diffraction spikes within the PC
point-spread-function.  The diffraction spike along the positive
direction of U3 axis (see HST WFPC2 Instrument Handbook 1995)
and in a direction $-$45$^\circ$ relative to the U3 axis are more
intense than those along the other two directions (the spike along the
$+$U3 axis being the most intense).  The model PSF does not
characterize this asymmetry accurately. On the other hand, while the
empirical PSF characterized this asymmetry well, because of the limited
number of standard star exposures available, the empirical PSF only
accurately characterized the PSF over a limited radius.  Hence the
empirical PSF was good for representing and removing the PSF structure
from the quasars that did not have highly saturated nuclei.  For the
quasar images that appeared to be only mildly saturated a small
residual sometimes was present along the brightest diffraction spike
(the spike along the +U3 direction), $\approx$1'' from the nucleus and
occupying a few pixels in diameter.  This distinct residual was easily
identifiable and removed using fits to the surrounding background. This
correction needed to be applied to about 12 of the quasar images and
those quasars are noted in Table 3.  This residual removal was carried
out mainly for cosmetic reasons.  The residuals contained very little
flux and since the residual was in every case well separated from the
quasar host, it has only a small effect on the final host morphology.
For four quasars (3C 205, 3C 279, 3C 351, and 3C 454.3) the nuclei were
highly saturated and we do not present the results of the PSF
subtraction due to their unreliability.  Generally, since
the exposure times for all the quasars were roughly the same, 
quasars with brighter nuclei were more difficult to determine
reliable host morphologies and brightnesses through PSF subtraction.

After PSF subtraction, the images were rotated so that north is at the
top of the frame and east is to the left.  Then each images was
smoothed with a 4 $\times$ 4 pixel median filter to remove the effects
of ``hot'' pixels and residual cosmic rays, to emphasize low surface
brightness features, and to reduce the additional noise in the final
image due to the noise in the image of the empirical PSF used for subtraction.
Contour plots of the final images are displayed in Figure 5 and 6.

We note that the relatively ``clean'' appearance of the contour plots
is due to the way in which they were constructed.  PSF subtracting the
images leads to an increase in the overall noise of the image near
the quasar.  After the image was smoothed by a 4 $\times$ 4 pixel median filter
we then selected the lowest contour to be at about the 3$\sigma$ noise
level in the region affected by the PSF subtraction, but well away from
the host galaxy.  Therefore, the minimum level is relatively high compared
to the noise level of the entire displayed image.  Picking such a
relatively high minimum contour level has the benefit of only showing
morphological features that have a high certainty of being real and not
artifacts of the PSF subtraction.  As noted above, any additional ``cleaning''
of the images was strictly limited to the removal of the residual along
the +U3 direction approximately 1'' from the nucleus in the quasars as noted
in Table 3.  Even this procedure had only a marginal influence on the
final displayed morphology.  In addition, in the one case (3C 179.0) where we
subtracted individually two images taken at different times, there
was close morphological agreement between the two images of the host galaxy
in spite of the rather dramatic change in the total magnitude of the
nucleus.

\subsection{On the Differences Between the EED Analysis and PSF Subtraction}

As can be seen in Table 3, there are rather large differences between
the amount of resolved fluxes estimated using the EED analysis and the
PSF subtraction.  These differences are not surprising.  First and
perhaps most importantly, the EED analysis is basically an integral
process and thus it is sensitive to low signal-to-noise, smoothly
distributed light, whereas the PSF subtraction analysis is inherently
differential and highlights the very small scale features lost in the
growth-curves.  Second, the EED analysis will always underestimate the
amount of resolved flux since it assumes that the fraction of host
light in the unresolved core (central $\sim$0.1 $\sq\arcsec$) is negligible
and then scales the contribution of ``fuzz'' under this assumption.
Thus one cannot simply measure the amount of flux from the central
$\sim$0.1 $\sq\arcsec$ to reconcile the estimates from the EED analysis with
that from the PSF analysis.  Moreover, it is known that as the focus of
the HST changes, different parts of the PSF are affected in different
ways (C.  Burrows and M.  McMaster, private communication).  Therefore,
using the diffraction spikes to gauge when to stop subtracting scaled
PSFs from the image, may not give the proper subtraction of the flux
from the nuclear region.  This was clearly evident for some of the
quasars where the flux would almost reach zero near the nucleus as
emission from the diffraction spikes disappeared into the noise.  Since
there is some disagreement in the amount of resolved flux between the
PSF subtraction and the EED analysis, in Figure 6, we provide images of
the quasars that are not classified as extended by the EED analysis,
but apparently have some extended flux in the PSF subtraction
analysis.  There are 8 such objects.

\subsection{On the Consistency of Our Results with Other Investigations}

Since it is difficult to ascertain robustly whether or not our procedure
for determining to what extent the quasars images are extended,
it is important to compare our results with those obtained
by other investigators.  Of course, this is challenging given the variety
of individual circumstances (HST versus ground-based data, use of adoptive
optics, different filters, optical versus IR images, etc) by which quasar
hosts have been observed.  Limiting ourselves to comparisons with other
HST programs to image quasar hosts in the optical, we can say that we find
broad consistency between the results presented here and those
of other programs investigating the hosts of radio loud quasars. For example, 
Boyce, Disney, \& Bleaken (1999) and Boyce et al. (1998), for small
samples of low-z radio-loud and radio-quiet quasars, found extended to
total flux ratios of roughly ten to a few tens of percent for the
radio-loud quasars in their studies which imply host magnitudes consistent
with our results.  At moderate redshifts, 0.4 $<$ z $<$ 0.5, in a study
of radio loud and radio-quiet quasar hosts, Hooper, Impey, \& Foltz (1997)
again found extended to total flux ratios of roughly ten to a few
tens of percent for their radio-loud subsample which also imply host magnitudes consistent with our results.  At the high redshifts (z$>$1), in two
small samples of radio-loud quasars, both Ridgway \& Stockton (1997; which
also included radio galaxies) and Lehnert et al. (1999)
find similar host galaxy magnitudes as we find here for similarly high
redshift quasars.

The broad agreement between the results of the study presented
here and those of other studies of radio-loud quasars using the WFPC2,
we are confident that our analysis is robust.  However, this statement
needs to be made more quantitative and we plan to make detailed comparisons
between our results and those of other studies in subsequent papers.

\section{The ``Alignment Effect'' for 3CR Quasars}

The ``Alignment Effect'' in which the axes of the optical and radio
emission roughly align is a well known effect observed in high redshift
radio galaxies (Chambers et al. 1987; McCarthy et al. 1987).  An
interesting test of various schemes attempting to relate the properties
of radio galaxies and quasars is to determine whether or not a sample
of quasars also exhibits a similar alignment.  To this end, we measured
the position angles of the radio images from the literature as given in
Table 2. These position angles were determined from the core of the
radio emission along the position angle of the jet.  If a jet was not
obvious in the radio image used, we then measured the position angle of
the highest surface brightness ``hot spot'' relative to the core.  For
the HST images, we measured the position angle by fitting a series of
ellipses as a function of surface brightness using the STSDAS program
``ellipse'' of the PSF subtracted, rotated and median smoothed quasar
images.  The position angle of the optical emission was taken to be the
position angle at a surface brightness of 21.5 m$_{F702W}$ arcsec$^{-2}$
for all of the quasars.  This value was chosen because it was the
surface brightness that was bright enough so that the ellipses that
were fitted to the data gave believable results with small
uncertainties for all of the quasar images (i.e., the uncertainty in the
ellipticity was $<$0.07 and the uncertainty in the PA was
$<$20$^\circ$).  We present the measured radio and optical position
angles in Tables 2 and 3 respectively and a histogram of the difference
in the radio and optical position angles in Figure 7.  Figure 7 shows a
tendency for the difference between the position angle of the principal
axes of the radio and optical emission to be less than 20$^\circ$.

We present this result for two reasons.  One of course is because this
is an important result for schemes that attempt to unify quasars and
radio galaxies based on viewing angle, evolution, or environment (e.g.,
Norman \& Miley 1984; Barthel 1989; Neff \& Hutchings 1990).  Our main
reason for presenting this result is that it lends support for our
methodology for determining if the quasar image is resolved and if so,
for determining the morphology of the underlying host galaxy.  Of
course it is important not to overstate this proposition.  When
comparing the ``alignment'' properties of radio galaxies and quasars
one obviously needs to worry about projections (especially for the
quasars), since the UV and radio are not likely to be perfectly aligned
in three dimensions.  In spite of this caveat, we would not expect to
see any correspondence between the radio and optical if the morphology
revealed through PSF subtraction happened by chance.  However, we
also note that there is a possible weakening of the alignment with
redshift and this may be an indication of the difficulty in resolving
high redshift hosts with short exposures through the WFPC2 (hence
we urge some due caution in over interpreting our results).  A more robust
analysis of the importance of this result and a comparison of the
strength of the ``alignment effect'' in quasars and radio galaxies will
be presented in a subsequent paper.

\section{Individual Source Descriptions}

In this section, we shall describe the morphologies of the individual
sources focusing on the following questions.  Are there any artifacts
in the images related to the PSF subtraction? What is the morphology of
the extended emission (isophotal size, ellipticity, orientation)?  Are
there signs of interactions with nearby companions?  How does the radio
structure relate to the optical morphology of the host as seen in the
HST data?  We shall also discuss the positional offsets between the
quasar nucleus and any nearby (in projection) companions seen in the
contour plots and the magnitudes of these companions.

\noindent
{\bf 3C 9, z=2.012}

3C 9 has the highest redshift of the quasars in our sample.  A short
(10 minute) ground-based U-band exposure of 3C 9 in Heckman et al.
(1991) did not reveal any extended structure.  Our HST observations 
do not detect extended flux.

\noindent
{\bf 3C 14, z=0.469}

The image of this source is approximately round.  At the faintest
isophotes it is preferentially extended along PA$\approx$120$^\circ$.
At higher isophotal levels, the orientation of the image is close to
PA$\approx$180$^\circ$.  There is a diffuse galaxy about 0.7''
south and 2.6'' west of the nucleus with a total magnitude of about
m$_{F702W}$ $\approx$ 23.2.  The radio source has a triple morphology
and is extended along PA$\approx$$-$5$^\circ$ and 170$^\circ$ (Akujor
et al. 1994) --- very similar to the PA of the highest surface brightness
extended optical emission.

\noindent
{\bf 3C 43, z=1.47}

3C 43 is optically one of the most compact objects in our sample.  The
PSF subtracted image has a faint extension along PAs$\approx$
30$^\circ$ and 180$^\circ$. Emission is extended to only about 1'' down
to a surface brightness of 22.9 m$_{F702W}$ arcsec$^{-2}$.  3C 43 has a
compact (LAS$\sim$2.6'') and complex radio structure (Sanghera et al.
1995; Akujor et al. 1991).  This complex morphology makes a direct
comparison between the optical and radio emission difficult, but it is
interesting to note that the ``{\bf U}-like'' structure seen in the
faintest isophote to the north of the nucleus is roughly mimicked in
the radio map of Sanghera et al. (1995).  The most northern component of
the complex radio structure has been identified as the nucleus by
Spencer et al. (1991).  If this is the case then the extended optical
emission to the south of the optical nucleus is roughly aligned with
the curved jet seen in the radio images of 3C 43 (e.g., Sanghera et al.
1995).   There is a nearby (in projection) companion galaxy about 3''
north and 0.2'' east of the quasar nucleus with a total magnitude,
m$_{F702W}$ $\approx$ 23.5.  There is another galaxy just visible on
the edge of the contour plot shown in Figure 5, which is almost
certainly a foreground galaxy.

\noindent
{\bf 3C 47, z=0.425}

3C 47 shows signs of interaction with a small galaxy approximately 1.7
arc seconds to the northeast of the nucleus.  There is a second galaxy
along this same direction approximately 3.5 arc seconds from the
nucleus.  These galaxies have total magnitudes of 21.6 m$_{F702W}$ and
21.7 m$_{F702W}$ respectively.  The elongated, off-center (i.e., not
centered on the nucleus) isophotes strongly suggest that the host
galaxy is interacting with one or both of the nearby (perhaps only in
projection) galaxies.  The 5 GHz radio map of Bridle et al. (1994) show
a core, jet, and two lobe morphology.  The jet is at
PA$\approx$210$^\circ$ which corresponds closely with a linear
feature seen in the HST image presented in Figure 5.

\noindent
{\bf 3C 68.1, z=1.238}

The HST data are consistent with a point source and thus we do not
detect any extended flux around 3C 68.1.

\noindent
{\bf 3C 93, z=0.357}

3C 93 possesses a host with a large angular size --- approximately 3''
in diameter (at a surface brightness of 22.9 m$_{F702W}$ arcsec$^{-2}$).
The isophotes are approximately round, with the brighter isophotes
being extended along the north south-direction and the faintest
isophotes oriented along PA$\approx$40$^\circ$.  The 1.5 and 8.4 GHz
radio maps of Bogers et al. (1994) show a ``core + double lobe'' radio
source with a relatively large angular size ($\sim$ 41'').  The radio
source is oriented along PA$\approx$45$^\circ$ and thus roughly
coincident with the principal axis of the outer isophotes.  The
morphology of the host galaxy agrees well with the R-band image of 3C
93 presented in Hes, Barthel, and Fosbury (1996).

\noindent {\bf 3C 138, z=0.759}

3C 138 is a flattened system oriented preferentially along
PA$\approx$130$^\circ$.  The inner isophotes become irregular with
extensions along PA$\approx$ 70$^\circ$ and along
PA$\approx$290$^\circ$.  3C 138 is also a so-called compact steep
spectrum (CSS) radio source.  The high resolution radio map of Redong et al.
(1991) shows a compact source whose linear and triplet structure is
extended on scales of a few tenths of an arc second along
PA$\approx$70$^\circ$.  This axis of emission is similar to the
extension of the isophotes on scales of a few tenths of an arc second
seen in the optical image.

\noindent
{\bf 3C 147, z=0.545}

The isophotes of 3C 147 are fairly flat in the high surface brightness
regions and become rounder at fainter isophotal levels.  The main axis
of the optical emission is at PA$\approx$55$^\circ$.  The galaxy is
about 2'' across down to about 22 m$_{F702W}$ arcsec$^{-2}$.  3C 147 is also a
well known compact steep spectrum radio source whose size is 0.5'' and
has a jet-like structure pointing out from the nucleus at
PA$\approx$240$^\circ$ (van Breugel et al. 1992).  Moreover, there is a
blob of emission 0.4'' from the nucleus at PA$\approx$25$^\circ$.  In
the HST image, we also see two blobs of emission about 1.4'' to the
south of 3C 147.  The total magnitude of these two blobs is about 21.7
m$_{F702W}$.  There is also another bright galaxy visible 3.3'' east of
the nucleus.  The total magnitude of this galaxy is 19.4 m$_{F702W}$.

\noindent
{\bf 3C 154, z=0.580}

The HST data are consistent with a point source in the EED analysis.
However, there is evidence for extended emission from the PSF
subtraction.  In Figure 6, we show the morphology of this extended
emission.  The F702W image of 3C 154 extended by about 1.5'' at a level
down to a surface brightness of 22 m$_{F702W}$ arcsec$^{-2}$.  Radio maps
of 3C 154 show a classical core, double lobe morphology (e.g., Bogers
et al. 1994).  The source is oriented along PA$\approx$100$^\circ$ and
is large (LAS$\sim$53''; Bogers et al. 1994).  The position angle of the
radio source corresponds to a faint extension in the HST image that
reaches about 1.2'' from the nucleus down to a surface brightness of 22
m$_{F702W}$ arcsec$^{-2}$.

\noindent
{\bf 3C 175, z=0.768}

The HST image of 3C 175 is marginally resolved.  The PSF subtracted
image shows a complex morphology.  The inner, brighter isophotes are
oriented preferentially east-west.  The fainter isophotes show a
``plume'' of emission to the south-east and south of the nucleus.  This
``plume'' reaches about 2'' from the nucleus (down to 22 m$_{F702W}$
arcsec$^{-2}$).  The radio source also has a ``classical'' radio structure
of core and two radio lobes.  This triple structure is oriented along
PA$\approx$240$^\circ$ (jet side).  The size of the radio source is
large --- LAS$\sim$56''.  In the optical image there is an extension in
the isophotes along PA$\approx$240$^\circ$.  There is no evidence from
ground-based optical images of 3C 175 that it is resolved (Malkan 1984;
Hes et al. 1996).

\noindent
{\bf 3C 179, z=0.856}

The images of 3C 179 were reduced using a different method than the
rest of the sample.  Two 300 second images were taken of 3C 179
separated by a period of a month.  Over that time, 3C 179 increased by
about 0.5 magnitudes in brightness.  We therefore PSF subtracted each
image individually and, aligned and rotated both residual images and
then averaged the two images.  The magnitude and fraction of the
extended emission was obtained by comparing the average quasar
brightness with that of the average host brightness.  Of course the
fraction of extended to total flux was different in the two images.

The brightest isophotes of the fuzz are oriented preferentially
in the east-west direction.  There is a bright ``knot'' of emission
about 0.8'' from the nucleus along a PA$\approx$270$^\circ$.  The
radio maps of Reid et al. (1995) show a complex morphology.  The
radio maps show a jet along PA$\approx$270$^\circ$ and a double lobe
morphology.

\noindent
{\bf 3C 181, z=1.382}

The image of 3C 181 does not appear to be spatially resolved in these
HST data.

\noindent
{\bf 3C 186, z=1.063}

The image of 3C 186 is not resolved according to the EED analysis.
However, the PSF subtraction suggests that it might be resolved.  We
show the possible morphology of the host in Figure 6.  The
image of 3C 186 shows ``fuzz'' about 1.8'' across down to a surface
brightness of 22 m$_{F702W}$ arcsec$^{-2}$.  There are two significant
position angles of extended emission, PA$\approx$30$^\circ$ and
PA$\approx$110$^\circ$.  There is a nearby (in projection) galaxy about
2.3'' from the nucleus along PA=65$^\circ$.  The total magnitude of
this nearby companion is m$_{F702W}$=22.2.  The radio morphology is
compact and is oriented along PA$\approx$140$^\circ$ (Rendong et al. 1991;
Spencer et al. 1991).

\noindent
{\bf 3C 190, z=1.195}

The PSF subtracted image of 3C 190 reveals a complex morphology.  The
principal axis of the optical emission is along
PA$\approx$140$^\circ$.  There is a clump of emission approximately
0.8'' to the east of the nucleus.  The radio morphology of 3C 190 is
compact (LAS $\sim$3'') and is linear having a chain of several hot
spots (Spencer et al. 1991) along PA$\approx$ 240$^\circ$.  In the HST
image, we see a distortion in the isophotes along the PA of the radio
structure.

\noindent
{\bf 3C 191, z=1.956}

There is no evidence for a resolved component in 3C 191 according to
the EED analysis,  but PSF subtraction analysis suggests that it is resolved.
The PSF subtracted image of 3C 191 reveals a complex morphology.
The general orientation of the fuzz is along PA$\approx$0$^\circ$ and
$\approx$200$^\circ$.  There is a ``plume'' of emission to the
southeast of the nucleus along PA$\sim$140$^\circ$.  The radio
morphology is a core plus double lobe morphology with a principal
axis of emission along PA$\approx$165$^\circ$ (Ankujor et al. 1991).

\noindent
{\bf 3C 204, z=1.112}

3C 204 has a relatively high percentage of extended to total emission
(almost 50\%).  The host galaxy is a flat system (e$\sim$0.25) and its
major axis is oriented preferentially along PA$\approx$150$^\circ$.
The radio emission has a core, jet, double lobe morphology with the jet
oriented along PA$\approx$275$^\circ$ (Reid et al. 1995).  There appears
to be a faint extension of emission along PA$\approx$275$^\circ$ in the
HST image of the host galaxy.  This ``finger'' of emission may be
related to the radio jet seen in the radio image of Reid et al. (1995).

\noindent
{\bf 3C 205, z=1.534}

The HST image of this quasar was saturated.  No PSF subtraction 
or EED analysis was attempted.

\noindent
{\bf 3C 207, z=0.684}

3C 207 does not appear to be extended in these HST data.

\noindent
{\bf 3C 208, z=1.110}

3C 208 does not appear to be extended in these HST data.

\noindent
{\bf 3C 215, z=0.412}

The optical counterpart of 3C 215 appears to have a flattened
elliptical structure with its major axis oriented along PA$\approx$
135$^\circ$.  The 5 GHz radio image of Bridle et al. (1994) shows a
complex structure with an inner region consisting of several high
surface brightness knots of emission along an approximately east-west
line, engulfed in a large, more diffuse emission oriented along
PA$\approx$150$^\circ$.  The radio emission is seen over a large scale,
LAS$\sim$ 1'.  A ground-based V-band image presented by Hes et al.
(1996) shows a similar morphology to the image presented here.

\noindent
{\bf 3C 216, z=0.67}

The HST image of 3C 216 indicates that the host galaxy is an
interacting system.  The faint isophotes of the quasar fuzz are not
centered on the quasar nucleus, but are offset to the northeast along
PA$\approx$30 --- 45$^\circ$.  The nearby (in projection) galaxy is about
1.6'' to the north of the nucleus and has a magnitude of 21.9.  There
is another brighter galaxy just off the edge of the contour plot
presented here is about 4'' to the east and 2.5'' north of the
nucleus.  The total magnitude of this galaxy is m$_{F702W} \approx$
20.4.  The 1.7 and 5 GHz radio images of Reid et al. (1995) show a
compact radio source (LAS $\sim$ 6'') oriented along
PA$\approx$40$^\circ$.  Along this PA lies a radio ``hotspot'' about
1'' from the nucleus.  This hotspot is approximately coincident with
the ``plume'' of optical emission we see in the HST image.

\noindent
{\bf 3C 220.2, z=1.157}

3C 220.2 does not appear to be extended in these HST data.

\noindent
{\bf 3C 249.1, z=0.313}

The HST image of 3C 249.1 is spectacular.  The extended emission
comprises about 70\% of the total light from the quasar.  A narrow-band
HST image centered on [OIII]$\lambda$5007 (Sparks, private comm.) shows
that most of the emission to the east of the nucleus is probably [OIII]
emission within the bandpass of the F702W filter (see also
Stockton \& MacKenty 1987).  However, the comparison with the
narrow-band [OIII] image suggests that much of the light from the inner
parts of the nebula is likely to be continuum emission from the host
galaxy.  A 5 GHz radio image of Bridle et al. (1994) reveals a core,
jet, double lobe morphology oriented preferentially along PA$\approx$
100$^\circ$.

\noindent
{\bf 3C 254, z=0.734}

The EED analysis of 3C 254 shows no evidence for resolution, but the PSF
subtraction indicates that it is probably resolved.  The HST image of 3C 254
has a complex morphology.  Its host galaxy is oriented approximately in
the east-west direction.  There are ``plumes'' of emission along
PA$\approx$45$^\circ$ and 285$^\circ$.  The second of these plumes
corresponds to the direction of the most distant radio lobe seen in the
5 GHz radio map of Reid et al. (1995).  This radio map reveals that 3C
254 has a double-lobed radio morphology with a central core.

\noindent
{\bf 3C 263, z=0.646}

The host galaxy of 3C 263 appears to be a flat system (e$\sim$0.3),
with its major axis aligned along PA$\approx$350$^\circ$.  There is a nearby
galaxy in projection along the major axis of the galaxy ($\approx$1.9''
from the nucleus) with a magnitude of 22.2.  There is also another
nearby galaxy about 0.2'' south and 1.5'' west of the nucleus.  This
galaxy has a total magnitude of about 22.3.  The 5 GHz radio map in
Bridle et al. (1994) shows a large scale (LAS$\sim$51'') core, jet,
double lobe source with the jet have an orientation of
$\approx$110$^\circ$.  The HST image shows a ``finger'' of extended
emission in the counter-jet direction (PA$\sim$300$^\circ$).

\noindent
{\bf 3C 268.4, z=1.400}

The image of 3C 268.4 does not appear to be extended in the EED
analysis.  PSF subtracting the image suggests that 3C 268.4 might be
extended.  In Figure 6, we show the possible resolved structure of the
quasar.  The HST image of 3C 268.4 reveals that it is another source
with a complex morphology.  The long axis ($\approx$1.5'') of the host
is approximately along PA = 230$^\circ$.  In the faintest isophotes
there is also a ``finger'' of emission pointing approximately to the
south.  In this direction there is a nearby (in projection) galaxy that
is about 2.6'' from the nucleus and has a total magnitude of 21.1.  The
1.4 and 5 GHz radio images of Reid et al. (1995) show a core, jet,
double lobe source with a largest angular size of about 12''.  The
principal axis of the radio emission is approximately along PA =
215$^\circ$ and corresponds roughly to the principal axis of the host
galaxy.

\noindent
{\bf 3C 270.1, z=1.519}

According to the EED analysis, 3C 270.1 does not appear to be
extended.  However, in Figure 6, we show the morphology of the possible
extended emission obtained by PSF subtraction which suggests that the
image of 3C 270.1 is resolved.  The host galaxy of 3C 270.1 is oriented
preferentially in the east-west direction (PA=100$^\circ$).  There are
``plumes'' of emission to the south and to the north-west.  The high
resolution radio image of Akujor et al. (1994) shows a compact
(LAS$\sim$10''), ``bent'', triple source (core $+$ two lobes) along
PA$\approx$180$^\circ$ and 320$^\circ$.

\noindent
{\bf 3C 277.1, z=0.321}

The host galaxy of 3C 277.1 is large compared to those in the rest of
the sample (emission is seen over 3'' down to 22 m$_{F702W}$
arcsec$^{-2}$).  The overall morphology of the host is round, but
distortions to the inner isophotes are seen in the directions of
PA$\approx$170$^\circ$ and 310$^\circ$.  The high resolution 1.7 and 5
GHz images of Reid et al. (1995) show a compact (LAS$\sim$1.5'') double
oriented along PA$\approx$310$^\circ$.  The position of the radio
``hotspot'' to the northwest roughly corresponds to the distortion we
see in the HST image of the host galaxy.

\noindent
{\bf 3C 279, z=0.538}

The image of this quasar was saturated.  No PSF subtraction was attempted.

\noindent
{\bf 3C 280.1, z=1.659}

The host galaxy of 3C 280.1 is compact and compared to most other hosts
in the sample, its surface brightness increases rapidly with decreasing
distance from the nucleus.  The isophotes are round (e$\sim$0.05) and
are oriented along PA$\approx$350$^\circ$.  The 5 GHz radio map of
Swarup, Sinha, \& Saika (1982; but see also Lonsdale et al. 1992 and
Akujor et al. 1994) shows a core, two hotspots (one close to the
nucleus, $\sim$ 1'' to the southeast, and a more distant one, about
12'' to the west-northwest) and then a wiggly chain of emission to the
southeast along PA$\approx$120$^\circ$.  This ``chain'' of radio
emission is seen from about 4'' to about 11'' from the nucleus.  The
host galaxy, as seen in our HST image, does have an outward bending of
the isophotes along the direction of and over the region of the
southeastern hotspot seen in the radio maps of Swarup et al. (1982).

\noindent {\bf 3C 287, z=1.055}

The HST data are not spatially resolved.

\noindent
{\bf 3C 288.1, z=0.961}

The HST data are not spatially resolved.

\noindent
{\bf 3C 298, z=1.436}

The host galaxy of 3C 298 is dominated by two morphological features ---
an ``arm'' of emission that projects from the nucleus to the south-west
(PA=225$^\circ$) and then bends around to the east and a ``plume'' of
emission to the north-northeast of the nucleus
(PA$\approx$20$^\circ$).  The radio images of Rendong et al. (1991) and
van Breugel et al. (1992) show a compact triple source (LAS$\sim$1.8'')
with an east-west orientation.

\noindent
{\bf 3C 309.1, z=0.905}

The host galaxy of 3C 309.1 is a flat elliptical galaxy.  The major
axis of the host galaxy is oriented along PA=130$^\circ$.  The high
resolution radio image of Redong et al. (1991) shows 3C 309.1 to be a
compact source, with a nuclear region oriented along
PA$\approx$145$^\circ$ and a LAS$\sim$0.1'' and a ``lobe'' about 1''
from the nucleus along PA$\approx$95$^\circ$.  The highest surface
brightness isophotes of the host have an orientation roughly like that
of the nuclear radio emission.

\noindent
{\bf 3C 334, z=0.555}

The HST image of 3C 334 shows a host galaxy that is distorted, having
twisted and off-center isophotes.  The general orientation of the host
is PA$\approx$120$^\circ$ and is 1.5'' across along its major axis
(down to a surface brightness of 21 m$_{F702W}$ arcsec$^{-2}$)..  The 5 GHz
radio image of Bridle et al. (1994) shows a large (LAS$\sim$57'') triple
(core, jet, two radio lobes) source.  The radio jet emerges from the
nucleus at PA$\approx$140$^\circ$ and curves around to the north.  The
orientation of the optical image of the host galaxy is approximately
the same as that of the radio (5 GHz) image.  Several ground-based
images of 3C 334 show that it is extended and has a morphology similar
to that presented here.  For example, an [OII]$\lambda$3727 image in
Hes et al. (1996) shows that 3C 334 is extended along
PA$\approx$10$^\circ$ and a [OIII]$\lambda$5007 image of Lawrence
(1990) shows extended line emission along PA$\approx$150$^\circ$.  Both
results have some morphological similarity to the image presented in
this study.

\noindent
{\bf 3C 343, z=0.988}

The image of 3C 343 is most unusual for the quasars imaged in this
sample; it did not require any PSF subtraction!  It appears to be a
flat system with a major axis of about 2'' long oriented along
PA$\approx$60$^\circ$.  3C 343 is another CSS radio source.  The radio
map of Rendong et al. (1991) reveals a compact (LAS $\approx$0.3'')
complex radio source.  There is a finger of emission (a jet?) pointing
out along PA$\approx$320$^\circ$.  From an analysis of an optical
spectrum, Aldcroft, Bechtold, \& Elvis (1994) suggest that 3C 343 is a
Seyfert 2 galaxy (i.e., the galaxy has narrow permitted and forbidden
lines).  Given the high radio luminosity of this galaxy, a more
appropriate classification is as a radio galaxy.  The characteristics
of the optical spectrum from Aldcroft et al. supports our imaging data
and our contention that 3C 343 does not appear to be a quasar.

\noindent
{\bf 3C 351.0, z=0.371}

The image of this quasar was saturated.  No PSF subtraction was attempted.

\noindent
{\bf 3C 380, z=0.692}

3C 380 appears to be marginally resolved and is perhaps an
interacting system.  It has two companion galaxies that appear to be
immersed in common isophotes with the host galaxy of 3C 380.  These
galaxies are approximately 0.6'' west and 0.5'' north and 0.8'' west
and 0.7'' north of the nucleus respectively.  The total magnitudes of
these two galaxy are m$_{F702W}$ $\approx$ 20.7  and m$_{F702W}$
$\approx$ 21.9.  The radio image of van Breugel et al. (1992) at 1.4 GHz
shows a very diffuse radio morphology that is strikingly similar but
larger than the optical HST image shown in Figure 5.  The interacting
companions are engulfed in this radio emission and there is a surface
brightness enhancement of the radio emission over the area of these
companions (Reid et al. 1995).

\noindent
{\bf 3C 418, z=1.686}

3C 418 appears to have a compact host galaxy with several nearby (in
projection) galaxies.   The long axis of the host is only
about 0.9'' across (down to a surface brightness of 22.9 m$_{F702W}$
arcsec$^{-2}$) and oriented preferentially along PA$\approx$225$^\circ$.
The two nearby galaxies are 0.9''W, 0.2''N and 1.2''W, 1.2''N from the
nucleus and have magnitudes of 23.8 and 23.0 m$_{F702W}$.  A 15 GHz
radio image of O'Dea, Barvainis, \& Challis (1988) shows a complex
radio morphology with several bends in a radio ``jet'' pointing
approximately along PA=330$^\circ$.  This twisted jet is approximately
2'' long.  The positions of the bends in the radio jet seem to
correspond roughly to the positions of these nearby ``companions'' seen
in the PSF subtracted HST image.  However, given the magnitudes of 
these ``companions'', it seems very unlikely that they would be at
the redshift of the quasar.

\noindent
{\bf 3C 432, z=1.785}

The HST data for 3C 432 are not extended according to the EED
analysis.  However, PSF subtraction suggests that 3C 432 may in fact be
resolved.  The HST image of 3C 432 shows a compact host about 1.2'' in
diameter down to a surface brightness of 22 m$_{F702W}$ arcsec$^{-2}$.  The
host is preferentially oriented along PA$\approx$ 45$^\circ$.  There is
a secondary ``plume'' of emission along PA$\approx$ 135$^\circ$.  A 4.9
GHz radio map of Bridle et al. (1994) shows a classical core, jet,
double lobed source extended over 15'' and oriented along
PA$\approx$135$^\circ$.

\noindent
{\bf 3C 454, z=1.757}

The HST image of 3C 454 reveals a round and compact ($\sim$1'' at a
surface brightness of 22 m$_{F702W}$ arcsec$^{-2}$) host galaxy.  The radio
morphology is also compact, with a largest angular size of about 0.8''
(Rendong et al. 1991;  Spencer et al. 1991).  The radio emission is
oriented preferentially in the north-south direction (Spencer et al.
1991).  As is the case for 3C 181 and 3C 288.1, we urge caution in
interpreting several of the features in the PSF subtracted image shown
in Figure 5.  Some of the structure seen along
PA$\approx$205$^\circ$ may be due to incomplete removal of the most
intense diffraction spike (i.e., the one in the +U3 direction).  The
questionable accuracy of the PSF removal is noted in Table 3.

\noindent
{\bf 3C 454.3, z=0.860}

The image of this quasar was saturated.  No PSF subtraction was attempted.

\noindent
{\bf 3C 455, z=0.5427}

The image of 3C 455 is interesting.  It has one of the highest
extended/total brightness ratios of the entire sample ($\sim$70\%).  It has
a simple diffuse morphology with an embedded high surface brightness
nucleus.  Down to a surface brightness of about 22 mag arcsec$^{-2}$, 3C
455 is about 2'' in diameter along its major axis at
PA$\approx$60$^\circ$.  The radio morphology in the 8 GHz map of Bogers
et al. (1994) is a core, double radio lobe type oriented along
PA$\approx$245$^\circ$ (very similar to the optical major axis seen in
the HST image).  The spectrum of this object shown in Gelderman \&
Whittle (1994) shows narrow permitted lines.  Combined with our result
that very little point-source subtraction was necessary, suggests that
this object should be re-classified as a compact radio galaxy rather
than a quasar.

\section{Concluding Remarks}

We can draw some general conclusions from an analysis of the images
presented in Figures 5 and 6.  We present HST ``snapshot'' images of a
sample of 43 quasars.  From a close inspection and analysis of these
data we draw the following conclusions:

\noindent
$\bullet$ Our analysis suggests that the quasar fuzz contributes from
$<$5\% to nearly 100\% in the most extreme case (about 20\% being
typical) of the total light from the quasar.  Although a large fraction
of the objects do not appear to be resolved in the EED analysis
($\sim$40\%), in about 1/2 of those sources, the more sensitive PSF
subtraction indicates the presence of a resolved component.

\noindent
$\bullet$ Many of the resolved sources show complex morphology with twisted,
asymmetric, and/or distorted isophotes and irregular extensions.

\noindent
$\bullet$ In almost every case of the quasars with spatially resolved
``fuzz'', there are similarities between the radio and optical
morphologies.  In many cases there are features with similar radio and
optical morphologies and/or the principal axes of radio and optical
(continuum and line) emission are roughly aligned.  This is further
evidence for the reality of the structures detected in the PSF
analysis.

\noindent
$\bullet$A significant fraction ($\sim$25\%) of sources show galaxies
nearby in project (within 5'') and some ($\sim$10\% of the sources)
show obvious signs of interactions with these nearby companions.

These results show that the generally complex morphologies of host
galaxies of quasars are influenced by the radio emitting plasma and by
the presence of nearby companions.

We must be cautious in interpreting these data since all of the
images have a contribution from emission lines to their morphologies
and brightnesses.  Separation of the continuum and line contributions
and constraints on the various mechanisms involved await new snapshot
surveys that are being carried out of 3C sources with the linear ramp
filters and another broad-band filter.  A more robust and quantitative
analysis of the data and the relationship between high redshift radio
galaxies and quasars will be presented in a future paper.

\acknowledgements
This research is partially supported by HST GO grant number
GO-5476.01-93A, by a program funded by the Dutch Organization for
Research (NWO) and by a NATO research grant.  M. D. L. would like to
thank Chris Burrows, Tim Heckman, and Matt McMaster for useful
suggestions.  We would also like to thank an anonymous referee
for her/his remarks that improved the presentation of our results.

\newpage

\centerline{FIGURE CAPTIONS}

\figcaption [] {The redshift distribution of the observed sample
of quasars from the 3CR catalogue.}

\figcaption [] {The 178 MHz radio power distribution of the
observed sample of 3CR quasars.}

\figcaption [] {The largest linear size distribution of the radio
emission for the observed sample of 3CR quasars.  We assumed
a cosmology of H$_0$=75 km s$^{-1}$ Mpc$^{-1}$ and q$_0$=0.5
in making this figure.}

\figcaption [] {The ratio of the QSO to PSF EEDs determined as
described in the text.  Those sources that have extended
fractions above 1.05 are considered resolved. The uncertainties
were determined assuming only that photon noise from
quasar and PSF light profiles and sky background contribute
to the uncertainties in the ratio
(i.e., we neglect possible systematic effects).}

\figcaption [] {a) Contour plots of PSF subtracted images of the quasars 3C
14 (z=1.469), 3C 43 (z=1.47), 3C 47 (z=0.425), and 3C 93 (z=0.357).
The lowest contours in each plot are:  22.3 m$_{F702W}$ arcsec$^{-2}$ for
3C 14, 22.9 m$_{F702W}$ arcsec$^{-2}$ for 3C 43, 22.9 m$_{F702W}$
arcsec$^{-2}$ for 3C 47, and 22.9 m$_{F702W}$ arcsec$^{-2}$ for 3C 93.  As
in every plot of this figure, each contour is an increase of a factor
of 2 (0.75 magnitudes) in surface brightness and each image has been
smoothed with a 4 pixel $\times$ 4 pixel median filter. b) Contour plots of PSF subtracted images of the
quasars 3C 138 (z=0.759), 3C 147 (z=0.545), 3C 175 (z=0.768), and 3C
179 (z=0.856).  The lowest contours in each plot are: 22.3 m$_{F702W}$
arcsec$^{-2}$ for 3C 138, 22.1 m$_{F702W}$ arcsec$^{-2}$ for 3C 147, 22.1
m$_{F702W}$ arcsec$^{-2}$ for 3C 175, and 22.9 m$_{F702W}$ arcsec$^{-2}$
\ for 3C 179. c) Contour plots of PSF subtracted images of the
quasars 3C 190 (z=1.195), 3C 204 (z=1.112), 3C 215 (z=0.412) and 3C 216
(z=0.67).  The lowest contours in each plot are:  22.7 m$_{F702W}$
arcsec$^{-2}$ for 3C 190, 22.7 m$_{F702W}$ arcsec$^{-2}$ for 3C 204, 21.9
m$_{F702W}$ arcsec$^{-2}$ for 3C 215, 22.1 m$_{F702W}$ arcsec$^{-2}$ for 3C
216. d) Contour plots of PSF subtracted images of the
quasars 3C 249.1 (z=0.313), 3C 263 (z=0.646), 3C 277.1 (z=0.321), and
3C 280.1 (z=1.659).  The lowest contours in each plot are:  21.7
m$_{F702W}$ arcsec$^{-2}$ for 3C 249.1, 20.6 m$_{F702W}$ arcsec$^{-2}$ for
3C 263, 22.1 m$_{F702W}$ arcsec$^{-2}$ for 3C 277.1, and 22.9 m$_{F702W}$
arcsec$^{-2}$ for 3C 280.1. e) Contour plots of PSF subtracted images of the
quasars 3C 298 (z=1.436), 3C 309.1 (z=0.905), 3C 334 (z=0.555), and 3C
343 (z=0.988).  The lowest contours in each plot are:  21.1 m$_{F702W}$
arcsec$^{-2}$ for 3C 298, 21.9 m$_{F702W}$ arcsec$^{-2}$ for 3C 309.1, 21.1
m$_{F702W}$ arcsec$^{-2}$ for 3C 334, 22.7 m$_{F702W}$ arcsec$^{-2}$ for 3C
343. f) Contour plots of PSF subtracted images of the
quasars 3C 380 (z=0.692), 3C 418 (z=1.686),  3C 454 (z=1.757), and 3C
455 (z=0.5427).  The lowest contours in each plot are:  21.9
m$_{F702W}$ arcsec$^{-2}$ for 3C 380, 22.9 m$_{F702W}$ arcsec$^{-2}$ for 3C
418, 22.3 m$_{F702W}$ arcsec$^{-2}$ for 3C 454 and 21.4 m$_{F702W}$
arcsec$^{-2}$ for 3C 455.}

\figcaption [] {a) Contour plots of PSF subtracted that are not resolved
according to the EED analysis (see text for details), but might be
resolved as shown by PSF subtraction.  The images of the quasars 3C 154
(z=0.58), 3C 186 (z=1.063), 3C 191 (z=1.956), and 3C 207 (z=0.684) are
shown.  The lowest contours in each plot are: 22.0 m$_{F702W}$
arcsec$^{-2}$ for 3C 154, 22.4 m$_{F702W}$ arcsec$^{-2}$ for 3C 186, 22.7
m$_{F702W}$ arcsec$^{-2}$ for 3C 191, and 21.9 m$_{F702W}$ arcsec$^{-2}$ for 3C
207.  As in every plot of this figure, each contour is an increase of a
factor of 2 (0.75 magnitudes) in surface brightness and each image has
been smoothed with a 4 pixel $\times$ 4 pixel median filter. b) Contour plots of PSF subtracted images of the
quasars 3C 254 (z=0.734), 3C 268.4 (z=1.4), 3C 270.1 (z=1.519) and 3C
432 (z=1.785).  The lowest contours in each plot are: 21.9 m$_{F702W}$
arcsec$^{-2}$ for 3C 254, 22.1 m$_{F702W}$ arcsec$^{-2}$ for 3C 268.4, 22.7
m$_{F702W}$ arcsec$^{-2}$ for 3C 270.1, and 22.4 m$_{F702W}$ arcsec$^{-2}$
\ for 3C 432.}

\figcaption [] {Histogram of the difference
between the position angles of the principal axis of the radio
and optical emission.}

\newpage

\begin{deluxetable}{rccccccc}
\tablecolumns{8}
\tablewidth{0pt}
\tablenum{1}
\tablecaption{Observation Log for 3CR Quasars}
\tablehead{
\colhead{3CR}&\colhead{N}&
\colhead{Int}&\colhead{Date}&
\colhead{3CR}&\colhead{N}&
\colhead{Int}&\colhead{Date} \\
\colhead{(1)}&\colhead{(2)}&
\colhead{(3)}&\colhead{(4)}&
\colhead{(5)}&\colhead{(6)}&
\colhead{(7)}&\colhead{(8)}}
\startdata
9.0&4&140&05/22/94&249.1&2&140&04/16/94 \nl
14.0&2&300&06/20/94&254.0&2&140&01/19/95 \nl
43.0&2&300&07/19/94&263.0&2&140&03/27/94 \nl
47.0&2&140&01/17/95&268.4&4&140&03/26/94 \nl
68.1&2&300&08/17/94&270.1&2&300&12/14/94 \nl
93.0&2&140&02/15/94&277.1&2&140&12/01/94 \nl
138.0&1&140&11/10/94&279.0&2&300&02/23/94 \nl
147.0&2&140&04/27/94&280.1&2&300&05/01/94 \nl
154.0&2&140&03/18/94&287.0&4&140&03/05/94 \nl
175.0&2&140&05/03/95&288.1&2&140&04/06/94 \nl
179.0&2&300&04/16/94&298.0&4&140&07/15/94 \nl
181.0&2&300&04/06/94&309.1&2&140&06/21/94 \nl
186.0&4&140&03/09/94&334.0&2&140&01/23/95 \nl
190.0&2&300&04/17/94&343.0&1&300&01/11/95 \nl
191.0&2&300&03/18/94&351.0&2&140&07/25/94 \nl
204.0&4&140&02/16/94&380.0&2&140&09/14/94 \nl
205.0&4&140&04/18/94&418.0&2&300&06/20/94 \nl
207.0&2&140&06/08/94&432.0&4&140&07/21/94 \nl
208.0&4&140&04/04/94&454.0&4&140&05/01/94 \nl
215.0&2&140&04/22/94&454.3&2&140&12/13/94 \nl
216.0&2&140&05/11/95&455.0&1&300&08/22/94 \nl
220.2&2&300&12/17/94&&&&
\enddata
\tablecomments{
Col. (1) and (5) --- 3CR source designation.
Col. (2) and (6) --- Number of separate exposures each with the
exposure time in listed in Col. (3) and (7) respectively.  The
total integration time is given by multiplying colums (2) and
(3) or colums (6) and (7).
Col. (3) and (7) --- Integration time in seconds per exposure.
Col. (4) and (8) --- Date of observation.  The two exposures for
3C179.0 and 3C279.0 were taken on different dates.  For 3C179.0,
the dates of the two exposures were 04/16/94 and 05/16/95,
for 3C279.0, the dates are 02/23/94 and 05/20/95.
}
\end{deluxetable}

\begin{deluxetable}{rcccccccccc}
\tablecolumns{11}
\tablewidth{0pt}
\tablenum{2}
\tablecaption{Basic Properties of the 3CR Snapshot Survey Quasars}
\tablehead{
\colhead{3CR}&\colhead{z}&
\colhead{V}&\colhead{S(178)}&
\colhead{log P$_{178}$}&\colhead{$\alpha$}&
\colhead{Class}&\colhead{LAS}&
\colhead{PA}&\colhead{LPS (kpc)}&
\colhead{ref} \\
\colhead{(1)}&\colhead{(2)}&
\colhead{(3)}&\colhead{(4)}&
\colhead{(5)}&\colhead{(6)}&
\colhead{(7)}&\colhead{(8)}&
\colhead{(9)}&\colhead{(10)}&
\colhead{(11)}}
\startdata
9&2.012&18.21&17.8&28.9&1.09&S&14&140&1.88& 4,5 \nl
14&1.469&20&10.4&28.4&0.81&S&26&355&2.17&1,2\nl
43&1.47&20&11.6&28.5&0.75&C&2.6&160&1.17&9,10\nl
47&0.425&18.1&26.4&27.9&0.98&S&85&215&2.57&4,6 \nl
68.1&1.238&19.5&12.8&28.4&0.80&S&53&175&2.48&4 \nl
93&0.358&18.09&14.4&27.5&0.82&S&41&40&2.22&3,6 \nl
138&0.759&17.9&22.2&28.2&0.46&C&0.6&70&0.51&6,8,12 \nl
147&0.545&16.9&60.5&28.4&0.46&C&0.8&240&0.59&3,12 \nl
154&0.5804&18&23.1&28.1&0.77&S&53&100&2.42&3 \nl
175&0.768&16.6&17.6&28.2&0.98&S&56&240&2.48&3,4 \nl
179&0.846&18.0&9.4&28.0&0.71&S&18&270&2.00&7 \nl
181&1.382&18.92&14.5&28.5&1.00&S&7.5&120&1.63&16 \nl
186&1.063&17.6&14.1&28.3&1.15&C&2.5&140&1.15&8,10 \nl
190&1.197&20&15.0&28.4&0.93&C&3&30&1.23&8,10,12 \nl
191&1.956&18.65&13.0&28.7&0.98&S&5.2&165&1.45&1,2 \nl
204&1.112&18.21&10.5&28.2&1.08&S&38&275&2.33&4,7 \nl
205&1.534&17.62&12.6&28.5&0.88&S&19&20&2.03&7 \nl
207&0.684&18.15&13.6&28.0&0.90&S&11&90&1.76&3 \nl
208&1.11&17.42&16.8&28.4&0.96&S&15&265&1.93&1,2,4 \nl
215&0.411&18.27&11.4&27.5&1.06&S&60&325&2.41&4,6 \nl
216&0.67&18.48&20.2&28.1&0.84&S&6&40&1.49&7 \nl
220.2&1.157&19&7.2&28.1&0.61&S&9&45&1.71&17 \nl
249.1&0.311&15.72&10.7&27.2&0.81&S&27&100&2.00&3,4 \nl
254&0.734&17.98&19.9&28.2&0.96&S&15&110&1.90&7 \nl
263&0.646&16.32&15.2&28.0&0.82&S&51&110&2.42&4 \nl
268.4&1.4&18.42&10.3&28.4&0.80&S&12&215&1.83&7 \nl
270.1&1.519&18.61&13.6&28.6&0.75&S&10&175&1.75&2 \nl
277.1&0.32&17.93&8.5&27.2&0.64&C&1.5&310&0.75&7 \nl
279&0.536&17.77&23.2&28.0&0.31&F&5&205&1.38&2,18 \nl
280.1&1.659&19.44&9.2&28.4&0.93&S&24&130&2.13&2,5,11 \nl
287&1.055&17.67&16.3&28.4&0.42&F&0.2&45&0.05&8,14 \nl
288.1&0.961&18.12&9.0&28.0&0.84&S&7&265&1.59&2 \nl
298&1.439&16.79&47.5&29.1&0.99&C&1.8&90&1.01&8,12 \nl
309.1&0.904&16.78&22.7&28.4&0.53&F&1.0&145&0.75&8 \nl
334&0.555&16.41&10.9&27.7&0.86&S&57&140&2.45&4 \nl
343&0.988&20.61&12.4&28.2&0.37&F&0.3&320&0.23&8 \nl
351&0.371&15.28&13.7&27.5&0.73&S&59&35&2.38&7 \nl
380&0.691&16.81&59.4&28.6&0.71&C&1.5&315&0.90&3,7,12 \nl
418&1.686&20&13.1&28.6&0.44&F&2.5&330&1.14&13 \nl
432&1.805&17.96&11.0&28.6&0.98&S&15&135&1.92&4 \nl
454&1.757&18.47&11.6&28.6&0.90&C&0.9&180&0.70&8,10 \nl
454.3&0.86&16.1&13.0&28.1&0.04&F&5&310&1.44&15 \nl
455&0.5427&19.7&12.8&27.8&0.71&C/S&4.4&245&1.33&3
\enddata
\tablecomments{
Col. (1) --- 3CR source designation.
Col. (2) --- Redshift of the source as taken from Spinrad et al. (1986).
Col. (3) --- The V magnitude of the quasar taken from Spinrad et al. (1986).
Col. (4) --- The 178 MHz flux density of the quasar
taken from Spinrad et al. (1986).
Col. (5) --- The logarithm of the 178 MHz power in Watts assuming
the cosmology q$_{0}$=0.5 and H$_0$=75 km s$^{-1}$ Mpc$^{-1}$.
Col. (6) --- The spectral index of the radio emission as taken from
Spinrad et al. (1986).
Col. (7) --- The classification of the radio source morphology and
spectral index.  Quasars classified as ``S'' are steep spectrum sources
with spectral indices greater than 0.7 (Col. 6) and largest projected
sizes greater than 20 kpc (Col. 10).  Quasars classified as ``C'' are compact
steep spectrum sources (CSS) or gigahertz peaked sources (GPS) and are
sources with spectral indices greater than 0.7 (Col. 6) and largest
projected sizes less than 20 kpc.  Quasars classified as ``F'' are
flat spectrum sources with spectral indices less than 0.7 (Col. 6).
Col. (8) --- The largest angular scale (LAS) in units of arc seconds
of the radio emission as
measured from the radio maps in the references given in Col. (11).
Col. (9) --- The position angle of the jet as measured relative to
the core measured from the radio maps in the references given in Col. (11).
In the few cases where a jet was not obvious, the position angle has
been measured from the core through the brightest hotspot.  The typical
uncertainty in this determination is about $\pm$10$^\circ$.
Col. (10) --- The largest projected physical size of the radio source
determined using the largest angular scale from Col. (8) and the
cosmology q$_0$=0.5 and H$_0$=75 km s$^{-1}$ Mpc$^{-1}$.  The unit is
the logarithm of the largest physical size in kpc.
Col. (11) --- The reference for the radio map used to measure the
properties tabulated in Cols. (8) and (9).  The references are 1=
Akujor et al. (1991), 2 = Akujor et al. (1994), 3 = Bogers et al. (1994),
4 = Bridle et al. (1994), 5 = Lonsdale et al. (1993), 6 = Price et al.
(1993), 7 = Reid et al. (1995), 8 = Rendong et al. (1991), 9 = Sanghera
et al. (1995), 10 = Spencer et al. (1991), 11 = Swarup et al. (1982), 12 =
van Breugel et al. (1992), 13 = O'Dea et al. (1988),14 = Fanti et al. (1989),
15 = Browne et al. (1982), 16 = Mantovani et al. (1994), 17 = Schilizzi,
Kapahi, \& Neff (1982), 18= de Pater \& Perley (1983).
}
\end{deluxetable}

\begin{deluxetable}{rccccccccc}
\tablecolumns{10}
\tablewidth{0pt}
\tablenum{3}
\tablecaption{HST Results on 3CR Snapshot Survey Quasars}
\tablehead{
\colhead{3CR}&\colhead{m$_{total}$}&
\colhead{${resolved \over total}$(EED)}&\colhead{${resolved \over total}$(PSF---)}&
\colhead{Rot}&\colhead{$\mu_{lim}$}&
\colhead{Resolved}&\colhead{PA}&
\colhead{O-R}&\colhead{Comments} \\
\colhead{(1)}&\colhead{(2)}&
\colhead{(3)}&\colhead{(4)}&
\colhead{(5)}&\colhead{(6)}&
\colhead{(7)}&\colhead{(8)}&
\colhead{(9)}&\colhead{(10)}}
\startdata
9&17.3&$<$0.05&...&...&...&N&...&?&... \nl
14&18.8&0.17&0.40&$-$65.0&22.3&Y&310&R&5 \nl
43&20.5&0.17&0.49&$-$66.7&22.9&Y&170&R?&6 \nl
47&17.7&0.17&0.38&113.6&22.9&Y&240&R&5 \nl
68.1&18.6&$<$0.05&...&...&...&N&...&?&... \nl
93&18.1&0.33&0.56&122.3&22.9&Y&170&R?&5 \nl
138&18.4&0.14&0.34&$-$39.0&22.3&Y&105&R&5 \nl
147&17.2&0.08&0.39&112.5&22.1&Y&250&R&5 \nl
154&16.9&$<$0.05&0.17&136.6&22.0&Y?&170&N&6 \nl
175&$<$16.4&0.05&$<$0.20&146.6&22.1&Y&230&R&1,3,7 \nl
179&18.5&0.12&0.38&136.7&22.9&Y&260&R&3,5 \nl
181&18.3&$<$0.05&...&...&...&N&...&?&... \nl
186&17.5&$<$0.05&0.32&155.8&22.4&Y?&140&R&5 \nl
190&18.9&0.07&0.28&146.4&22.7&Y&90&R?&6 \nl
191&17.8&$<$0.05&0.36&140.7&22.7&Y?&210&N&3,5 \nl
204&17.8&0.05&0.18&$-$159.2&22.7&Y&330&N&5 \nl
205&...&...&...&...&...&N&...&?&2 \nl
207&17.5&$<$0.05&0.34&153.6&21.9&Y?&90&R&3,5 \nl
208&18.2&$<$0.05&...&...&...&N&...&?&... \nl
215&17.7&0.14&0.40&151.6&21.9&Y&310&R&5 \nl
216&18.7&0.25&0.59&147.5&22.1&Y&30&R&5 \nl
220.2&$<$17.7&$<$0.05&...&...&...&N&...&?&... \nl
249.1&$<$15.8&0.20&$<$0.73&$-$173.0&21.7&Y&120&R?&1,3,5 \nl
254&17.4&$<$0.05&0.32&$-$57.1&21.9&Y?&100&R&5 \nl
263&15.9&0.14&0.11&$-$146.8&20.6&Y&160&N&3,7 \nl
268.4&17.3&$<$0.05&0.29&$-$139.6&22.1&Y?&230&R&5 \nl
270.1&$<$18.0&$<$0.05&$<$0.35&$-$18.5&22.7&Y?&90&N&1,3,5 \nl
277.1&17.6&0.27&0.56&$-$3.9&22.1&Y&310&R&5 \nl
279&...&...&...&...&...&N&...&?&2 \nl
280.1&18.4&0.12&0.41&$-$150.4&22.9&Y&175&N&3,5 \nl
287&17.5&$<$0.05&...&...&...&N&...&?&... \nl
288.1&17.4&$<$0.05&...&...&...&N&...&?&... \nl
298&$<$16.4&0.10&$<$0.18&155.4&21.1&Y&135&N&1,6 \nl
309.1&17.1&0.07&0.32&$-$176.6&21.9&Y&120&N&3,5 \nl
334&$<$16.4&0.13&$<$0.21&$-$15.9&21.1&Y&140&R&1,3,6 \nl
343&20.0&0.81&1.00&4.3&22.7&Y&80&R&5 \nl
351&...&...&...&...&...&N&...&?&2 \nl
380&16.9&0.05&0.36&148.5&21.9&Y&310&R&3,5 \nl
418&19.6&0.07&0.43&$-$98.6&22.9&Y&45&N&6 \nl
432&17.8&$<$0.05&0.33&$-$106.4&22.4&Y?&200&N&3,5 \nl
454&18.4&0.18&0.49&$-$50.6&22.3&Y&140&N&4(206),6 \nl
454.3&...&...&...&...&...&N&...&?&2 \nl
455&18.8&0.21&0.69&$-$109.2&21.4&Y&240&R&5 \nl
\enddata
\end{deluxetable}

Col. (1) --- 3CR source designation.
Col. (2) --- Total magnitude within an aperture of radius
1.4 arc seconds.
Col. (3) --- The resolved fraction of the total emission
within a 1.4 arc second aperture measured using the EED analysis.
See text for details.
Col. (4) --- The resolved fraction of the total emission within a 1.4
arc second aperture measured by conducting PSF subtraction.  See text
for details.  For those sources where the nucleus is saturated, we have
given the measured resolved fraction as an upper limit.
Col. (5) --- Angle (in degrees) that the image was rotated to make
north at the top and east to the left in each image.  Positive values
imply a counter-clockwise direction of the rotation.  To calculate the
PA of the brightest diffraction spike one uses the following formula:
If the rotation angle is positive, then one subtracts 45$^\circ$ from
the listed rotation angle. If the rotation angle
is negative, then one adds 315$^\circ$ to the listed rotation angle.  Possible
residuals due to the diffraction spikes are noted in individual source
descriptions in \S 7 and in Col. (10) of this table.
Col. (6) --- Surface brightness limit of the lowest contour of the
plots shown in Figure 5 and 6 in units of magnitudes arcsec$^{-2}$.
Col. (7) --- Is the quasar image resolved?  ``Y'' implies that both
the EED and PSF subtraction analyses suggest that the quasar image
is resolved.  ``Y?'' implies that only the PSF subtraction analysis
suggests that the quasar image is resolved.  ``N'' implies that the
image is not resolved according to both analyses.
Col. (8) --- Position angle of the principal axis of the host galaxy
measured north through east.  The typical uncertainty
is this determination is about $\pm$20$^\circ$ (see \S6 for details).
Col. (9) --- Is there an association between the radio and optical
morphologies?  A ``R'' implies that (i) the principal axis of the radio
emission is within 20$^\circ$ of the principal axis of the optical
emission or (ii) that there is a detailed correspondence between the
morphological features seen in both the radio and optical. ``R?''
implies that there is some similarity in the optical and radio
morphologies which may indicate a relationship between the radio and
optical emission.  ``N'' implies that there is little or no correspondence
between the radio and optical emission.
Col. (10) --- Comments. A ``1'' implies that the nucleus is saturated
and that PSF subtraction was attempted.  A ``2'' implies that the
nucleus was very saturated and that PSF subtraction was not attempted
due to the severe saturation of the nucleus and the large angular
scales over which the diffraction spikes are seen due to the nuclear
point source.  A ``3'' implies that the image has had a residual of the
diffraction spike in the ``+U3'' direction removed by fitting a surface
to the surrounding background (see text for details).  A ``4'' implies
that there is a possible residual in the displayed image due to the
diffraction spike in the ``+U3'' direction.  The number in parentheses
after the ``4'' is the approximate position angle (measured relative to
north through east) of the possible residual.  A ``5'', ``6'', or ``7''
is used to characterize the level of uncertainty in the magnitude of
the host galaxy due to the PSF subtraction.  A ``5'' implies that the
uncertainty in the magnitude of the host galaxy is about $\pm$0.2
magnitudes.  A ``6'' implies that the uncertainty in the magnitude of
the host galaxy is about $\pm$0.4 magnitudes.  A ``7'' implies that the
uncertainty in the magnitude of the host galaxy is about $\pm$0.7
magnitudes.

\end{document}